\documentstyle[twocolumn,aps]{revtex}
\begin{document}
\draft
\twocolumn[\hsize\textwidth\columnwidth\hsize\csname@twocolumnfalse\endcsname
%\preprint{HEP/123-qed}
\title{Breakdown of Modulational Approximations in Nonlinear Wave Interaction}
\author{G.J.L. Gerhardt, M. Frichembruder, F.B. Rizzato\footnote{
e-mail:rizzato@if.ufrgs.br; tel.:+55 51 3166470; fax:+55 51 3191762}}
\address{Instituto de F\'{\i}sica - Universidade Federal do Rio Grande 
do Sul, Caixa Postal 15051, \\ 91501-970 Porto Alegre, Rio Grande do 
Sul, Brazil}
\author{S.R. Lopes}
\address{Institute for Plasma Research, University of Maryland, \\ 
College Park, Maryland 20742, USA \\
and \\
Departamento de F\'{\i}sica - 
Universidade Federal do Paran\'a, Caixa Postal 19081, \\ 81531-990 Curitiba, 
Paran\'a, Brazil}
\maketitle
\date{\today}
\begin{abstract}

In this work we investigate the validity limits of the modulational 
approximation as a method to describe the nonlinear interaction of 
conservative wave fields. We focus on a nonlinear Klein-Gordon equation and 
suggest that the breakdown of the approximation is accompanied by a transition 
to regimes of spatiotemporal chaos.
  
\end{abstract}
]
%****************************************************************************  
%.......................................
%****************************************
%
\section{Introduction}

\noindent Modulational instability of high-frequency nonlinear 
waves is a common process in a variety of circumstances involving 
wave propagation in continuous systems. Modulational processes can be seen 
to occur in a wide range of physical situations, from nonlinear waves in 
plasmas \cite{dter78} to nonlinear electromagnetic waves propagating in optical 
fibers \cite{mal97}. What usually happens in all those cases is that due to 
generic nonlinear interactions, the amplitude of a high-frequency carrier 
develops slow modulations in space and time. If the modulations are indeed much 
slower than the high-frequencies involved, one can obtain simplified equations 
describing the dynamics of the slowly varying amplitudes solely, the amplitude 
equations \cite{lili91,novo99,olig96,riz98,erich98}. In the present analysis 
we consider systems that become integrable in this modulational limit, a 
feature often displayed. Should this be indeed the case, 
no spatiotemporal chaos would be observed there. The basic interest 
then would be to see what happens when the approximations leading to 
modulational approximations cease to be satisfied. The paper is organized as 
follows: in \S 2 we introduce our model equation and discuss 
how and when it can be approximated by appropriate 
amplitude equations; in \S 3 we investigate the modulational process from the 
point of view of nonlinear dynamics; in \S4 we perform full spatiotemporal 
simulations and compare the results with those obtained in \S 3; and in \S 5 
we conclude the work. 

\section{Model equation, modulational approximations, and amplitude equations}

In the present paper we 
focus attention on a nonlinear variant of the Klein-Gordon equation (NKGE) to 
investigate the breakdown of modulational approximations in the context of 
nonlinear wave fields. The NKGE used here reads
\begin{equation}
\partial_t^2 A(x,t) - \partial_x^2 A(x,t) + {\partial \Phi \over \partial A}=0,
\label{eqzero}
\end{equation}
($\partial_t \equiv \partial / \partial t, \> 
\partial_x \equiv \partial / \partial x$) 
where we write the generalized nonlinear potential as 
\begin{equation}
\Phi(A) = \omega^2 \> {A(x,t)^2 \over 2} - {A(x,t)^4 \over 4} 
+ {A(x,t)^6 \over 6},
\label{phi}
\end{equation}
$\omega$ playing the role of a linear frequency which will set the 
fast time scale. The remaining coefficients on the right-hand-side 
were chosen to 
allow for modulational instability and saturation; we shall see that 
while the negative sign of the second fulfills the condition for 
modulational instability, the positive sign of the third provides saturation. 
The choice of their numerical values is arbitrary, but our results are 
nevertheless generic. The NKGE is known to describe wave propagation in 
nonlinear media and the idea here is to see how the dynamics changes as a 
function of the parameters of the theory: wave amplitude, and time and length 
scales. 

Let us first derive the conditions for slow modulations. We start by supposing 
that the field $A(x,t)$ be expressed in the form 
\begin{equation}
A(x,t) = \tilde A (x,t) e^{i \omega t} + {\rm complex\>\>conjugate}. 
\label{fatora}
\end{equation}
Then, if one assumes slow modulations and discards terms like 
$\partial_t^2 \tilde A$ and the highest-order power of the potential 
$\Phi$, one obtains 
\begin{equation}
2 i \omega \partial_t \tilde A(x,t) - \partial_x^2 \tilde A(x,t) - 
3 |\tilde A(x,t)|^2 \tilde A(x,t) = 0, 
\label{schroclassica}
\end{equation}
which is, apart from some rescalings, the Nonlinear Schr\"odinger Equation - 
we shall refer to it as NLSE here - an integrable equation. 

This is no novelty; it is known that the modulational approximation is 
obtainable 
when there is a great disparity between the time scales of the high-frequency 
$\omega$ and the modulational frequency, we call it $\Omega$, such that 
terms of order $\Omega^2 \tilde A$, when compared to $\omega^2 \tilde A$, can be 
dropped from the governing equation. The magnitude of the modulational frequency 
can be estimated as follows. Consider Eq. (\ref{schroclassica}) and suppose a 
democratic balance among the magnitude of its various terms; 
$\omega \partial_t \tilde A \sim \partial_x^2 \tilde A \sim {\tilde A}^3$. Then one 
obtains for $\partial_t \rightarrow \Omega$,
\begin{equation}
{\Omega \over \omega} \sim ({\tilde A \over \omega})^2.
\label{condition}
\end{equation}
It is thus clear that the modulational approximation is valid only 
when $\tilde A \ll \omega$, since this condition slows down the modulational 
process causing $\Omega \ll \omega$. The next question would be on what is to 
be expected when the modulational approach ceases to be valid. Before 
proceeding along this line, let us mention that a stability analysis 
can be performed on  Eq. (\ref{schroclassica}). One perturbs an homogeneous 
self-sustained state with small fluctuations of a given wavevector $k>0$ (we 
choose $k>0$ here, but the theory is invariant when $k \rightarrow - k$) and 
after some algebra one concludes that: (i) the field in the homogeneous state, 
let us call this field $A_h$, is given by 
\begin{equation}
A_h = a_o e^{- i {3 a_o^2 \over 2 \omega} t}
\label{comp1}
\end{equation}
where $a_o$ is an arbitrary 
amplitude parameter and where the exponential term should be seen as providing a 
small nonlinear correction to the linear frequency $\omega$; and (ii) the 
perturbation is unstable when 
\begin{equation}
k < k_{tr} \equiv \sqrt{6} a_o.
\label{modinst}
\end{equation}
with maximum growth rate at
\begin{equation}
k_{max} \equiv {k_{tr} \over \sqrt{2}}.
\label{maxgrowth}
\end{equation}
When unstable, the homogeneous state typically evolves towards a state 
populated by regular structures which can be formed precisely because 
the underlying governing NLSE, Eq. (\ref{schroclassica}), is of the 
integrable type as mentioned earlier. 

\section{Beyond the modulational approximation}

To advance the analysis beyond the modulational regimes we start from 
the basic equation, Eq. (\ref{eqzero}), but do not use approximation 
$\partial_t^2 \tilde A \ll \omega^2 \tilde A$ leading to Eq. (\ref{schroclassica}) 
and eventually to condition $\tilde A \ll \omega$. The idea is precisely to 
examine what happens as the ratio $\tilde A / \omega$ of Eq. (\ref{condition}) 
grows from values much smaller, up to values comparable to the unit.

Our first task is to examine how the regular and well known modulational 
instability analyzed in the previous section comes directly from Eq. (\ref{eqzero}). 
To do that, let us write a truncated solution as the sum of an 
homogeneous term plus fluctuations with wavevector $k$, 
$A = A_h(t) + A_1(t) (e^{i k x} + e^{-i k x})$, $A_h$ and $A_1$ real. The 
truncation, that discards higher harmonics, is legitimate within linear 
regimes, but fortunately we shall see that it is not as restrictive as it might 
appear even in nonlinear regimes. The general idea favouring truncation here 
is that in the reasonable situation where modes with the fastest growth rates 
are more strongly excited, relation (\ref{maxgrowth}) indicates that second 
harmonics are already outside the instability band since 
$2 k_{max} = \sqrt{2} k_{tr} > k_{tr}$. Under these circumstance 
one would be led to think that the most important modes would be the 
homogeneous and those at the fundamental spatial harmonic. We shall actually 
see that the truncation provides a nice and representative approach to the 
case of regular regimes. 

Note that we are already considering the amplitudes of the exponential functions as 
equal. This results from a simplified symmetrical choice of initial conditions and is 
totally consistent with the real character of Eq. (\ref{eqzero}). After some lengthy 
algebra, one finds out that the coupled nonlinear dynamics of the fields $A_h$ and 
$A_1$ is governed by the Hamiltonian
$$
H={p_h^2 \over 2} + {\omega^2 q_h^2 \over 2} - {q_h^4 \over 2} + 
{2 q_h^6 \over 3} + 
{p_1^2 \over 2} + {\chi^2 q_1^2 \over 2} - {3 q_1^4 \over 4} + 
{5 q_1^6 \over 3} - 
$$
\begin{equation}
3 q_h^2 q_1^2 + 10 q_h^4 q_1^2 + 15 q_h^2 q_1^4,
\label{hamiltonian}
\end{equation}
where $\chi^2 \equiv \omega^2+k^2$, $q_h = A_h/\sqrt{2}$, $q_1 = A_1$, and 
where the $p$'s denote the two momenta conjugate to the respective 
$q$-coordinates.

The Hamiltonian (\ref{hamiltonian}) can be informative. As a first 
instance it can be used to determine the stability properties of the homogeneous 
pump, as mentioned before. To see this, assume that in average 
$q_h \gg q_1$ and solve the dynamics perturbatively. In zeroth order, 
one would have the following Hamiltonian governing the $(p_h,q_h)$ 
dynamics:
\begin{equation}
h_o = {p_h^2 \over 2} + {\omega^2 q_h^2 \over 2} - {q_h^4 \over 2} + 
{2 q_h^6 \over 3}.
\label{ho}
\end{equation}
With help of action-angle variables ($\rho,\Theta$) for the 
zeroth-order part and conventional perturbative techniques 
the solution reads 
\begin{equation}
q_h = \sqrt{2 \rho \over \omega} \cos(\omega t - {3 \over 2} 
{\rho \over \omega^2} t), 
\label{ordemzero}
\end{equation}
if $\rho$, the amplitude parameter, is not too large. Note that the 
oscillatory frequency undergoes a small nonlinear correction which 
is determined by the quartic term in $q_h$ of the Hamiltonian $h_o$. 

Next we consider the driven Hamiltonian controlling the dynamics of 
the canonical pair $(p_1,q_1)$
\begin{equation}
h_1 = {p_1^2 \over 2} + {\chi^2 q_1^2 \over 2} + 3\>q_h^2 q_1^2,
\label{h1}
\end{equation}
where we recall that the pair $(p_1,q_1)$ describes the inhomogeneity 
of the system. One again introduces action-angle variables $(I,\theta)$ 
to rewrite the linear Hamiltonian (\ref{h1}) in the form
\begin{equation}
h_1 = \chi I + 12 {\rho \over \omega} {I \over \chi} \cos^2 \theta 
\cos^2 (\omega t - {3 \over 2} {\rho \over \omega^2} t),
\label{h11}
\end{equation}
from which we obtain the resonant form
\begin{equation}
h_{1,r} = {1 \over 2 \omega^2}(k^2 - 3\>\rho) I + 
{3\>\rho I \over 2 \omega^2} \cos (2 \phi),
\label{hres}
\end{equation}
with $\phi = \theta -(\omega - 3\rho / 2\omega^2)t$, and where use is made of the 
approximation $\chi \approx \omega + k^2/(2 \omega)$ valid when 
$k \ll \omega$. Now consider the stability of a 
small perturbation $I \sim 0$. For this initial condition 
$h_{1,r} \rightarrow 0$. 
Since $h_{1,r}$ is a constant of motion, solutions for arbitrarily large 
values of $I$, what would indicate instability, are possible only 
when $|\cos(2 \phi)| \le 1$. 
From the resonant Hamiltonian (\ref{hres}), 
this demands 
\begin{equation}
k < \sqrt{6 \rho \over \omega},
\label{cond2}
\end{equation}
and also indicates that maximum growth rate for $I$ occurs at 
$k_{max}=\sqrt{3 \rho / \omega}$. Comparisons of temporal dependence 
of Eqs. (\ref{fatora}), (\ref{comp1}) and (\ref{ordemzero}) shows that 
$\rho / \omega = a_o^2$, and that conditions (\ref{modinst}) and (\ref{cond2}) 
are therefore one and the same as they should be. In other words, starting 
from the full nonlinear wave equation, we recover the typical results 
naturally yielded by the NLSE.

But the Hamiltonian (\ref{hamiltonian}) gives further information because 
apart the truncation, it is not an adiabatic approximation like 
Eq. (\ref{schroclassica}); it can therefore tell us whether the reduced 
dynamics, if unstable, is of the regular or chaotic type. The interest on this 
issue comes from the fact that the reduced dynamics usually helps to determine 
the spatiotemporal patterns of the full system: while regular reduced dynamics 
is associated with regular structures - frequently a collection of 
solitons or soliton-like structures, chaotic reduced 
dynamics is associated with spatiotemporal chaos. The correlation has its roots 
on the so called stochastic pump model \cite{lili91}. The model states that 
intense chaos in a low-dimensional subsystem of a multidimensional environment can 
make the subsystem act like a thermal source, irreversibly delivering energy into 
others degrees-of-freedom. In the limit of a predominantly regular dynamics 
undergoing in the reduced system, 
irreversibility is greatly reduced and energy tends to remain confined 
within the subsystem. On the other hand, in the limit of deeply chaotic 
dynamics with no periodicity at all, energy flow out of the subsystem is fast 
and there is not even much sense in defining the subsystem as an approximately 
isolated entity. The intermediary cases are those more amenable to a 
description in terms of the stochastic pump \cite{lili91,novo99}. We point out, 
however, that in all cases, even in the chaotic one, analysis of the reduced 
subsystem serves as an orientation on what to be expected in the full 
spatiotemporal dynamics. In our case, the appropriate subsystem is precisely 
the one we have been using. This is so because it is the smallest subsystem 
comprising the most important ingredients of the dynamics: the homogeneous 
and the only linearly unstable modes.

\begin{figure}[h]
\vspace{4in}
\includegraphics{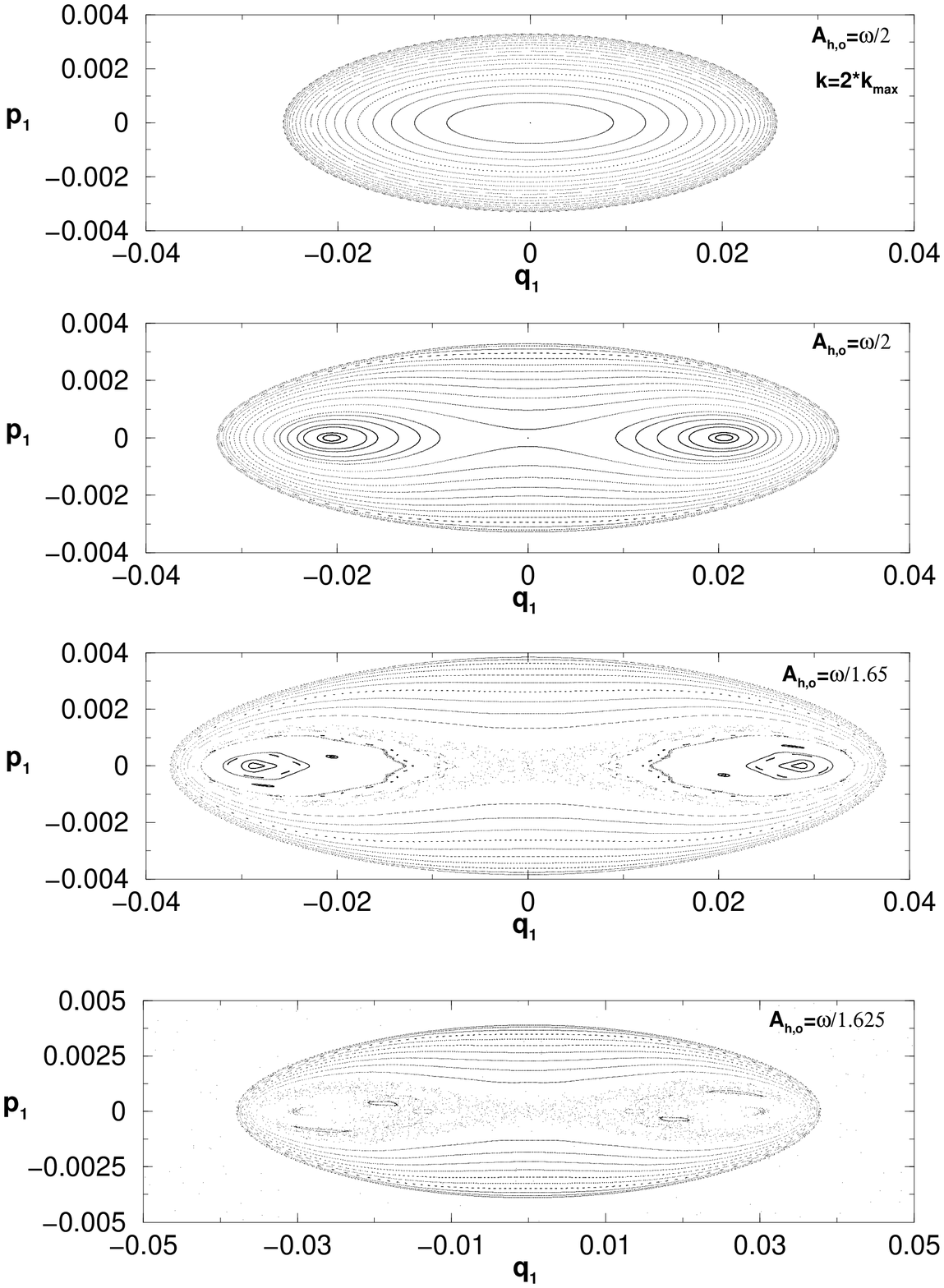}
\caption{Low-dimensional Poincar\'e plots on the projected
phase-space $(p_1,q_1)$. $\omega=0.1$ in all cases, and $k=k_{max}$ in panels 
(b)-(d). Initial conditions discussed in the text.}
\label{fig1.ps}
\end{figure}

We now examine these points in some detail. Let us consider $\omega = 0.1$. When
$A_{h,o} \ll \omega$, $A_{h,o} \equiv A_h(t=0)$, the modulational
approximation can be used to determine the instability range, $k < k_{tr}$. 
A surface of section plot based on the two-degrees-of-freedom 
Hamiltonian (\ref{hamiltonian}) produces Fig. (\ref{fig1.ps}), where we record 
the values of $(q_1,p_1)$ whenever $p_h = 0$ with $dp_h/dt > 0$, and 
where we take $q_1 = 0.01 q_h \ll q_h$ and $p_h = p_1 = 0$ to determine the 
total unique energy of the various initial conditions we launch in the 
simulations. Note that the particular ``seed'' initial condition introduced 
above is included in the ensemble of initial conditions launched and represents 
small perturbations to an homogeneous background.
In Fig. (\ref{fig1.ps}a) 
we take $A_{h,o}/\omega = 0.5$ and $k = 2 k_{max} > k_{tr}$. One lies outside 
the instability range and the figure reveals that the origin, which 
represents the purely homogeneous state $q_1=A_1=0$, is indeed stable. 
The remaining panels are all made for $k=k_{max}$ and increasing values of 
the ratio  $A_{h,o}/\omega$. One sees that for such values of $k$ and 
$A_{h,o}$ not only the origin is rendered unstable, as it also becomes 
progressively surrounded by chaotic activity. The inner chaotic trajectory 
issuing from the origin - i.e., the trajectory representing the 
modulational instability - is encircled by invariant curves, but the last panel 
already shows that some external chaos, here represented by scattered points 
around the invariant curves, is also present. At some amplitude 
$A_{h,o} = A_{critical}$ slightly larger than the one used in panel (d), invariant 
curves are completely destroyed. Above the critical field, inner orbits are no 
longer restricted to move within confined regions of phase-space - these 
orbits are in fact engulfed by the external chaos seen in panel (d). External 
chaos here is a result of the hyperbolic point positioned at the local 
maximum of the generalized potential $\Phi(A)$, at $A_1 \sim 0.1$.
One gross determination of the critical field based on simple 
numerical observation yields $A_{critical}/\omega \sim 1/1.62$, although 
the corresponding transition from order to disorder in the full simulations 
may not be so sharply defined. Yet, one may expect that in both 
low-dimensional and full simulations, the transition should occur at 
comparable amplitudes.    

\section{Full spatiotemporal simulations}

\begin{figure}[h]
\vspace{4.25in}
\includegraphics{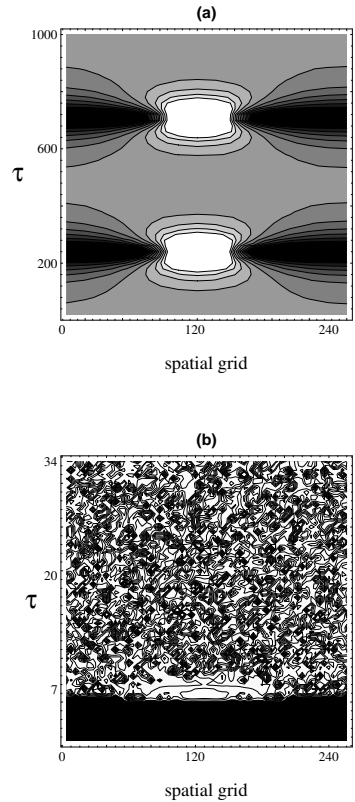}
\caption{Full spatio-temporal simulations for $\omega = 0.1$ and $k=k_{max}$.
$A_{h,o}/\omega=0.1$ in (a) and $A_{h,o}/\omega=1/1.625$ in (b).
The intensity plots are made for the quantity
$Q = \sqrt{\omega^2 A(x,t)^2 + {\dot A}(x,t)^2}$; lighter shades are associated
with larger values of $Q$, which varies within the range $0 < Q < 0.0014$ in
(a) and $0 < Q < 0.8$ in (b). $\tau \equiv \omega t / 2 \pi$ here and in all
the remaining simulations}
\label{fig2.ps}
\end{figure}

To make the appropriate comparisons, we now look into the full simulation of 
Eq. (\ref{eqzero}). Full simulations are made through a discretization of the 
spatial domain via a finite difference method. The dynamics is resolved temporally 
by means of a sympletic integrator, as was the purely temporal Hamiltonian. The 
results are quite robust and energy is conserved up to $10^{-6}$ parts in one. 
In Fig. (\ref{fig2.ps}) we display the spacetime history of the quantity 
$Q = \sqrt{\omega^2 A(x,t)^2 + \dot A(x,t)^2}$ - we plot this quantity 
because it becomes a constant of motion in the limit where we discard nonlinearities 
and inhomogeneities. Initial conditions are the same as the seed initial 
condition used in Fig. (\ref{fig1.ps}), but the control parameters differ. In 
the case of Fig. (\ref{fig2.ps}a) where we choose $A_{h,o}/\omega = 0.1$ so as 
to safely satisfy $A_{h,o} \ll A_{critical}$ and $A_{h,o} \ll \omega$, 
it is seen that the 
spatiotemporal dynamics is very regular. The homogeneous state is unstable, but 
only periodic spatiotemporal spikes can be devised. This is the regular 
spatiotemporal dynamics so typical of the integrable NLSE and this kind of 
dynamics agrees very well with the low-dimensional predictions of the reduced 
Hamiltonian (\ref{hamiltonian}). The fact that only one single structure can be 
found along the spatial axis at any given time, indicates that the dynamics is 
singly periodic (period-1) along this axis and thus basically understood in 
terms of the reduced number of active modes (homogeneous plus fundamental 
harmonics) of Hamiltonian (\ref{hamiltonian}). We now move into the vicinity of 
the critical amplitude $A_{critical}$ discussed earlier. Under such conditions, 
one may expect to see the effects of spatiotemporal chaos. The value 
$A_{h,o}/\omega = 1/1.625$ is chosen in Fig. (\ref{fig2.ps}b), where full 
simulations indeed display a highly disorganized state after a short regular 
transient. Regularly interspersed 
spikes can no longer be seen and spatial and temporal periodicities are lost, 
which characterizes spatiotemporal chaos. In this regime many modes become 
active (we shall return to this point later) and the reduced Hamiltonian 
(\ref{hamiltonian}) fails to provide an accurate description of the full 
dynamics. However it still provides a good estimate on the point of transition. 
A more throughful examination of the transition in the full simulations 
suggests that the critical field there is a bit smaller - a value close to 
$\omega / 1.95$; for smaller values we have not observed noticeable 
signals of spatiotemporal chaos even for much longer runs than those 
presented here. The much larger oscillations 
executed by $Q$ in chaotic cases (see the legend of Fig. (\ref{fig2.ps})) 
is a direct result of the destruction of the invariant curves seen in 
Fig. (\ref{fig1.ps}). In the absence of invariant curves, initial conditions 
are no longer restricted to move near the origin.  
It is thus seen that there are limits to an integrable modulational 
description of the dynamics of a wave field. The limits are essentially 
set by the parameter $A_{h,o} / \omega$. If it is much smaller than 
the unit, the modulational description is valid and one can 
expect to see a collection of spatiotemporal periodic structures 
being formed as asymptotic states of the dynamics. On the other hand, as the 
parameter approaches the unit, nonintegrable features are likely to be 
seen. In particular, regularity does not survive for very long, and 
fluctuations with various length scales appear in the system. This 
is a regime of spatiotemporal chaos which fundamentally involves the 
presence of nonlinear resonances between the frequency of the 
carrier, $\omega$, and the intrinsic nonlinear modulational frequency, 
$\Omega$.

\begin{figure}[b]
\vspace{4.25in}
\includegraphics{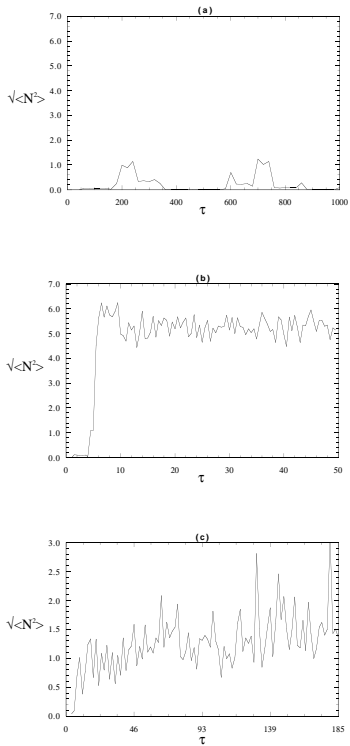}
\caption{Average number of active modes as a function of time. Parameters
respectively equal to those of the previous figure with exception of
panel (d) where we take $A_{h,o}/\omega = 1/1.95$.}
\label{fig3.ps}
\end{figure}

Due to the presence of chaos, one is suggested that the transition involves an 
irreversible energy flow out of the reduced subsystem. If one computes the average 
number of active modes
\begin{equation}
<N^2> \equiv {\sum_n n^2 |A_n|^2 \over \sum_n |A_n|^2},
\label{soma}
\end{equation}
where the amplitudes are defined in the form
\begin{equation}
A_n = \sum_j A(x_j,t) e^{i n k x_j},
\label{aene}
\end{equation}
one obtains the plots shown in Fig. (\ref{fig3.ps}). 
\noindent In Eq. (\ref{aene}), 
\lq\lq$j$'' is the discretization index. In Fig. (\ref{fig3.ps}a) we use the 
same conditions as in Fig. (\ref{fig2.ps}a). This panel 
shows that in rough terms, energy keeps periodically migrating between the 
homogeneous mode (when $\sqrt{<N^2>} \sim 0$) and the fundamental harmonic 
(when $\sqrt{<N^2>} \sim 1$). Conditions of Fig. (\ref{fig3.ps}b) are the same 
as those of Fig. (\ref{fig2.ps}b); one sees that as the ratio 
$A_{h,o}/\omega$ grows, periodicity is lost, and that energy flow out of the 
initial subsystem into other modes becomes clearly irreversible. In 
Fig. (\ref{fig3.ps}c) we use the same previous conditions with exception 
of $A_{h,o}$ which we take $A_{h,o} = \omega / 1.95$. This slightly smaller, 
but not too small, value of the initial amplitude allows to observe the 
slow diffusive transit of energy during the initial stages of the 
corresponding simulations. During this stage one can actually look at 
the subsystem as an energy source adiabatically delivering energy into other 
modes - the concept of the stochastic pump applies more appropriately in those 
situations.

\section{Final conclusions}

To summarize, in this paper we have studied the breakdown of modulational 
approximations in nonlinear wave interactions. We have analyzed 
a nonlinear Klein-Gordon equation to draw the following 
conclusions. Adiabatic or modulational approximations are accurate while the 
high-frequency of the carrier wave keeps much larger than the modulational 
frequency. Under these circumstances the 
full spatiotemporal patterns are regular as is the dynamics 
in the reduced subsystem where the system energy is initially injected. There 
is no net flow of energy out of the reduced subsystem into the remaining 
modes.

On the other hand, 
when both frequencies become of the same order of magnitude, the reduced 
subsystem undergoes a transition to chaos. Correspondingly, the spatiotemporal 
patterns of the full system become highly disordered and energy spreads 
out over many modes. The correlation between the low-dimensional and high-dimensional 
spatiotemporal chaos has its roots on the stochastic pump model \cite{lili91}. 
According to the model, a low-dimensional chaotic subsystem may act like a thermal 
source, delivering energy in a irreversible fashion to other degrees-of-freedom 
of the entire system. Spectral simulations performed here indicates 
that this seems to be the case with the present setting. 

The transition to 
chaos involves a noticeable increase in terms of wave amplitude. This 
takes place when the invariant curves of Fig. (\ref{fig1.ps}) are destroyed, 
allowing for the merge of the external and internal chaotic bands into one 
extended chaotic sea. 
This merging of chaotic bands is actually a result of reconnections involving the 
manifolds of the unstable fixed point at the origin, and other hyperbolic points 
associated with the curvature of the generalized potential $\Phi$ \cite{corso98}. 
When both manifolds reconnect, inner trajectories issuing from the origin start to 
execute the large and irregular oscillations seen in Fig. (\ref{fig2.ps}). More 
detailed studied of the process is under current analysis.  

\acknowledgments
This work was partially supported by 
Financiadora de Estudos e Projetos (FINEP), Conselho Nacional de 
Desenvolvimento Cient\'{\i}fico e Tecnol\'ogico (CNPq), and 
Funda\c{c}\~ao da Universidade Federal do Paran\'a - FUNPAR, Brazil. S.R. Lopes 
wishes to express his thanks for the hospitality at the Plasma Research 
Laboratory, University of Maryland. Part of the numerical work was performed on 
the CRAY Y-MP2E at the Supercomputing Center of the Universidade Federal do 
Rio Grande do Sul.  
%

%*********************************************
\newpage

%\begin{figure}[h]
%\vspace{3.25in}
%\special{psfile=fig1.eps angle=-0 voffset=-5 hoffset=20 hscale=30 vscale=30}
%\caption{Low-dimensional Poincar\'e plots on the projected
%phase-space $(p_1,q_1)$. $\omega=0.1$ in all cases, and $k=k_{max}$ in panels 
%(b)-(d). Initial conditions discussed in the text.}
%\label{fig1.ps}
%\end{figure}
%
%\begin{figure}[h]
%\vspace{3.25in}
%\special{psfile=fig2.eps angle=-0 voffset=-180 hoffset=0 hscale=50 vscale=50}
%\caption{Full spatio-temporal simulations for $\omega = 0.1$ and $k=k_{max}$. 
%$A_{h,o}/\omega=0.1$ in (a) and $A_{h,o}/\omega=1.625$ in (b).
%The intensity plot are made for the quantity 
%$Q = \sqrt{\omega^2 A(x,t)^2 + {\dot A}(x,t)^2}$; lighter shades are associated 
%with larger values of $Q$, which varies within the range $0 < Q < 0.0014$ in 
%(a) and $0 < Q < 0.8$ in (b). $\tau \equiv \omega t / 2 \pi$ here and in all 
%the remaining simulations.}
%\label{fig2.ps}
%\end{figure}
%
%\begin{figure}[h]
%%\vspace{2.25in}
%%\special{psfile=fig3.eps angle=-0 voffset=-80 hoffset=0 hscale=37 vscale=37}
%\caption{Number of active modes as a function of time. Parameters 
%respectively equal to those of the previous figure with exception of 
%panel (d) where we take $A_{h,o}/\omega = 1/1.95$.}
%\label{fig3.ps}
%\end{figure}
%
%

\begin{references}
%\begin{description}
%
\bibitem{dter78}S.G. Thornhill and D. ter Haar, Phys. Reports
{\bf 43}, 43 (1978).
%
\bibitem{mal97}B. Malomed, D. Anderson, M. Lisak, and M.L. Quiroga-Teixeiro, Phys. 
Rev. E {\bf 55} 962 (1997).
%
\bibitem{lili91}A.J. Lichtenberg and M.A. Lieberman, 
{\it Regular and Chaotic Motion}, Springer (1991).
%
\bibitem{novo99}S.R. Lopes and F.B. Rizzato, Phys. Rev. E {\it Nonintegrable 
Dynamics of the Triplet-Triplet Spatio-Temporal Interaction} submitted (1999).
%
\bibitem{olig96}G.I. de Oliveira, L.P.L. de Oliveira, and F.B. Rizzato, 
Phys. Rev. E {\bf 54}, 3239 (1996).
%
\bibitem{riz98}F.B. Rizzato, G.I. de Oliveira, and R. Erichsen, 
Phys. Rev. E {\bf 57}, 2776 (1998).
%
\bibitem{erich98}R. Erichsen, G.I. de Oliveira, and F.B. Rizzato, 
Phys. Rev. E {\bf 58}, 7812 (1998).
%
\bibitem{lop98}S.R. Lopes and F.B. Rizzato, Physica D {\bf 117}, 13 (1998).
%
\bibitem{corso98}G. Corso and F.B. Rizzato, Phys. Rev. E {\bf 58}, 
8013 (1998).
%
\end{references}
\end{document}